\documentstyle[prl,aps]{revtex}
\begin{document}
\title{Hydrogen Bonds in Polymer Folding}
\author{Jesper Borg, Mogens H. Jensen, Kim Sneppen and Guido Tiana}
\address{Niels Bohr Institute and Nordita, Blegdamsvej 17, DK-2100 {\O},
Denmark}
\maketitle


\begin{abstract} 
The thermodynamics of a homopolymeric chain with both Van der Waals
and highly--directional hydrogen bond interaction is studied. The
effect of hydrogen bonds is to reduce dramatically the entropy of
low--lying states and to give raise to long--range order and to
conformations displaying secondary structures. For compact polymers a transition
is found between helix--rich states and low--entropy sheet--dominated states.
The consequences of this transition for protein folding
and, in particular, for the problem of prions are discussed.
\vspace{0.5cm}


\end{abstract} 

\vskip0.5cm


Secondary structures are a prime feature of all structured polymers
and proteins \cite{Pauling,Alberts}. 
These structures are stabilized by
hydrogen bonding. For example, $\alpha$--helices of proteins are known
to be stabilized by hydrogen bonds which involve couples of donor and 
acceptor atoms belonging to consecutive turns of the helix \cite{Alberts}. 
Similarly, hydrogen bonds are responsible for
stabilizing the helix conformation in the
helix--coil transition in amino acid homopolymers \cite{Scheraga}.
Recently, there has been a wide developement of simplified lattice
models for protein folding, where each monomer interacts 
with its neighbours through
an isotropic Van der Waals interaction \cite{mj}.
These studies, in general, show no evidence of secondary
structure formation for chains of realistic length
 and  mostly deal with statistical features of good folders \cite{Derrida,Bryngelson}.

In this paper we discuss the effect of including directed
hydrogen bindings in simplified polymer models.
As a starting point we
adopt a lattice implementation widely used in the literature 
\cite{flory,go75,lau,sh_begin}. 
This model is defined by a string of monomers 
placed on subsequent positions of a 3-d cubic lattice.
The energy of a configuration $\{ {\bf r}_i \}$ 
is defined by the Hamiltonian

\begin{equation}
H_{VW} \; = \; - \; \sum_{i<j} \; \epsilon_{ij} \; 
\delta(|{\bf r}_i-{\bf r}_j|-1)
\end{equation}
where $\sum$ include all monomer pairs, 
the $\delta$ function ensures contributions
from nearest neighbours only and $\epsilon_{ij}$
contains the strength of 
Van der Waals interactions. In the case of homopolymers, all 
$\epsilon_{ij}$ are equal, while for proteins different
choices can be adopted \cite{lau,mj,Tang97}.

In addition to $H_{VW}$, we introduce in the Hamiltonian a
term associated with a directed interaction between couples of monomers.
On each monomer $i$ we assign a spin ${\bf s}_i$ 
representing a hydrogen donor--acceptor pair. 
This can be easily pictured as a spin because of the 
opposite directions of the $O$ (H acceptor)
and the $N$ (H donor) atom on the peptide backbone.
The spin is constrained to be perpendicular to the backbone.
To study the secondary structure of proteins, donors and acceptors coming from 
the amino acid sidechains should be taken into account as well. 
In this simple model we 
consider the minimal scenario of a homopolymer
with only one acceptor and one donor for each monomer.
The hydrogen bond part of the Hamiltonian reads

\begin{equation}
H_{H} \; = \; - \; 
\epsilon_H \; \sum_{ij} \; 
\delta( {\bf s}_i \cdot {\bf s}_j \; -\; 1) \;
\delta(|{\bf r}_i-{\bf r}_j| \; -\; 1),
\end{equation}
where only ${\bf s}_i$ that are perpendicular to the backbone are allowed,
and thus interactions along the backbone are automatically ignored.

Setting $\epsilon_{ij}=\epsilon_V$ for all $i$ and $j$, the Hamiltonian 
$H(\epsilon_V,\epsilon_H) = H_{VW}+H_H$
specifies the energy of any homopolymer configuration
including hydrogen bindings.
To study the equilibrium properties we 
sample configurations of the chain using the
simulated tempering techniques suggested by 
Marinari and Parisi \cite{Marinari}
and developed in the context of proteins
by Irb\"ack and Potthast \cite{Irback}. 
This method consists in examining an ensemble of different temperatures, 
by sampling 
a generalized partition function that includes the temperature as a 
dynamical variable. By adjusting the sampling rate associated with each 
temperature, one avoids trapping in local minima. 
The sampling rates are adiabatically adjusted according 
to the multiple histogram equations \cite{Ferrenberg} and, to avoid
overestimation of metastable states, 
we test for thermodynamic compatibility as described in ref. \cite{Borg}.

Figure 1 shows examples of structures obtained at low temperatures
using $H(0,2)$ and $H(1,2)$,
respectively. In the first case, one can observe helix--like structures
that, however, differ from true $\alpha$ helices by the change in
chirality along the helix. This is due to lattice constraints \cite{Borg}.
We denote this kind of structures pseudo--helices (``p--helix''),  
and quantify them for a given polymer conformation
by counting the number of bonds involved in them.

For example, in Fig. 1(a) the  monomers between 2 and 7 initiate
a pseudo helix, where neighbours 2--5 and neighbours 4--7 contribute.
The p--helix continues until monomer 22, which breaks it because this monomer
is not neighbour to any members of a p--helix.
A new pseudo--helix is initiated at monomer 25 and last throughout the chain.
A helix is similarly defined as a structure built as that of Fig. 1(a),
but with constant chirality (it would have been the case if the position of
monomer 7 and 9 were interchanged in Fig. 1(a) ).

Fig. 1(b) displays the ground state of a polymer
where Van der Waals interactions are included. In this case
one can observe structures resembling $\beta$--sheets.
The sheets can be either parallel
or antiparallel, as in natural proteins, 
in both cases quantified by identifying at least three
pairs of consequtive neighbours in a line 
(that is for parallel sheet it would be $\{(i,j),(i+1,j+1),(i+2,j+2)\}$ and
for antiparallel sheet $\{(i,j),(i+1,j-1),(i+2,j-2)\}$). 
In Fig. 1(b), for example, monomer pairs
$(1,10)$, $(2,9)$, $(3,8)$ and $(4,7)$ contribute to an antiparallel sheet, 
that gets broken at monomer 10. Monomers 6-13 
also participate in an antiparallel sheet with the layer above.

Both the folds displayed in Fig. 1 reveal long range order.
In particular, Fig. 1(b) displays an up--down symmetry and an organization where
sheets are on one side of the structure and the backbone connections 
between layers are concentrated on the opposite side.
The key--result of this analysis is, in fact,
that spin interactions induce large scale organization even
in the case of a homopolymer collapse.

Furthermore, the presence of hydrogen bonds causes a dramatic reduction of
entropy, compared to that found for homopolymers
with only isotropic interactions. The conformation displayed in Fig. 1(b)
is, in fact, the {\it unique}, zero entropy ground state for a 36mer
interacting through the Hamiltonian $H(1,2)$
(except for trivial symmetries, i.e. lattice symmetries 
and flipping of lines of spins. The latter being easily removed by
introducing a diedral term in the Hamiltonian). 
The reduction in entropy can be appreciated from the inset to Fig. 2.

Fig. 2 also displays the number of bonds involved in the four types
of secondary structures ($I$), as function of total number of bonds ($N_B$).
The dependence with number of bonds is obtained by thermal averaging
as function of temperature. The choice of using $N_B$ rather than the 
temperature as free variable is more convenient, since we are
comparing systems with different energy scales.
The three curves represent the case where only Van der Waals energy
is present ($\nu\equiv \epsilon_H/\epsilon_V =0$, full line),
where Van der Waal and spin coupling energy 
are present ($\nu=2$), and the case where
only spin coupling is present ($\nu=\infty$). 
For any degree of compactness the amount of secondary structure increases
with increasing spin coupling. In general, it also increases with compactness,
however a backbending for the $\nu=2$ case is present at nearly 
maximal compactness.
The backbending for $\nu=2$ in both plots is the mark of a phase transition.
The transition takes place at almost maximum compactness 
(which is associated with low temperature), 
where entropy is reduced abruptly while compactness changes to a minor extent.
In fact, this transition distinguishes between two types of compact polymers, 
a phase in which p--helices are predominant, 
and an ordered, highly symmetric one, rich in $\beta$-sheets (cf. Figs. 1b,3).

Fig. 3 shows in detail how the different types of secondary
structures change with respect to temperature $T=1/\beta$. 
The backbending in Fig. 2 corresponds to the transition at 
$\beta=3/\epsilon_V$, in which relatively disordered states with a large
fraction of pseudo--helices are replaced  
with ordered, sheet--dominated structures like that displayed in Fig. 1b.
We notice that even the conformations at intermediate values of the temperature
($\beta$ in the range between $1/\epsilon_V$ and $2.5/\epsilon_V$) are quite
ordered, in the sense that they have a large degree of helix structures. 
This behaviour can be compared with the case of homopolymers
without spin interactions, which have significantly less structures, and
in particular have much lesser helix content (data not shown).
The conformations mentioned above 
are rather compact, the homopolymer collapse taking place at $\beta \sim
1/\epsilon_V$, while the helix--sheet transition takes place at $\beta \sim
3/\epsilon_V$. The gap between the two transitions are adjustable by changing
$\nu$, f.ex. for $\nu=3$ we find the two transitions much closer to each other.

Most interestingly, the energy landscape of even the simple homopolymer turns
out to be extremly rough, so rough that the helix--sheet transition cannot be 
sampled with any normal Metropolis approach.
In other words, a polymer with hydrogen bonds displays a population
of low energy conformations which are
structurally very different from each other, separated by high energy
barriers and resembling rather closely the spin glass behaviour \cite{parisi}.

It is remarkable that the present model predicts a large variety
of helix structures, contrasting to a few, highly--ordered sheet structures.
This fact suggests that helix--like conformations are entropically favoured
in the protein folding mechanism, and consequently can act as intermediates
leading the chain to its equilibrium state, while 
sheet--like structures first appear
late in the folding process and contribute to the stabilization of the
ground state. A similar pattern is found
in the study of prion diseases \cite{prion}, 
where the sane, helix--dominated native form of the 
protein seems to be 
a very long-lived metastable easily accessible state, whereas the true
ground state is dominated by $\beta$--sheets and prone to aggregation.

Finally, we would like to stress that the current parametrization 
of hydrogen bonding is the simplest possible one. The driving force in
protein folding is believed to be hydrophobicity, where non--polar 
amino acids are deficient in hydrogen bonds 
and thereby cause an ordered, entropically unfavourable 
arrangement of the hydrogen bonds in the surrounding water. 
On the contrary, hydrophilic amino acids contain hydrogen receptors and donors
that can replace those that water cannot build due to the presence of the 
protein surface.
Thus, a more realistic model should explicitly include water.
To model protein realistically, one should also consider chains composed of monomers 
with different amount of hydrogen donors and acceptors,
reflecting the different types of amino acids.
In such a way one will be able to control the sequence of secondary structures
in a more specific way than with heterogeneous Van der Waals forces alone.
\vspace{0.5cm}

In summa, we have implemented a minimal model to keep into account hydrogen
bond effects in polymer folding. Already at the level
of homopolymers we have observed pronounced secondary structures,
structures which are ubiquitous in the realm of natural proteins.

\newpage

\begin{figure}

\caption{
(a) A typical low energy structure for the homopolymer
when only spin coupling is included.
The different gray scales of the monomers are only meant to help the
visualization of the 3-D structure. 
(b) The ground state when both spin and isotropic 
Van der Waals interactions are included ($\nu=2$).}
\end{figure}

\begin{figure}
\caption{
Total number of nearest neighbours, $I$, involved in the four types of structures,
as function of total number of nearest neighbours $N_b$ (i.e., as function of compactness).
One can observe that the more hydrogen bonding (i.e. higher
$\nu=\epsilon_H/\epsilon_V$) the more secondary structure is present
and, as revealed by the inset, the less entropy $\Delta S=S(T)-S(T=\infty)$ is
displayed by the system.  Furthermore, one notices that a too large hydrogen
bond energy prevents the formation of a compact state, and thereby of an
ordered ground state. At intermediate values of $\nu$, there appears a highly
ordered ground state that is sharply separated from a disordered compact state by a structural transition.}
\end{figure}

\begin{figure}
\caption{Number of bonds involved in various structures, $N_i$, as function 
of $\beta=1/T$ for $\nu=2$. The curves refer to antiparallel sheets
(``a--sheet'', solid line), pseudo--helices (``p--helix'', dashed line),
helices (``helix'', dot--dashed line) and parallel sheets (``sheet'', dotted lines). 
One can observe two transitions, one at low 
$T$, associated with the melting of the sheet--dominated structure (cf. Fig
1(b) ), and one at high $T$ (i.e. low $\beta$), associated to the collapse
transition of the polymer. At intermediate $T$ helices are important, serving
as intermediates for polymer folding, by reducing the temperature.}
\end{figure}



\begin{thebibliography}{99}

\bibitem{Pauling}

L. Pauling, R.B. Corey and H.R. Branson,
PNAS {\bf 37} 1951, 729-740.

\bibitem{Alberts} B. Alberts et. al.,

``Molecular Biology of the Cell'',Garland Publishing, New York 1994.

\bibitem{Scheraga}

H.A. Scheraga, {\sl Pure and applied Chemistry} {\bf 36} (1973) 1.

\bibitem{mj} S. Miyazawa and R. Jernigan, Macromolecules, {\bf 18}, 534 (1985)

\bibitem{flory}

P. J. Flory, J. Chem. Phys {\bf 17}, 303 (1949)

\bibitem{go75} 

N. Go, Int. J. Peptide Prot. Res. {\bf 7}, 313 (1975)

\bibitem{lau}

K. F. Lau and K. Dill, Macromolecules {\bf 22}, 3986 (1989)

\bibitem{sh_begin}

E. I. Shakhnovich and A. M. Gutin, J. Chem. Phys., {\bf 93}, 5967 (1989)

\bibitem{Derrida}

B. Derrida, Phys. Rev. Lett. {\bf 45} (1980) 79 

\bibitem{Bryngelson}

J. Bryngelson et. al. Proteins {\bf 21} (1995) 167.

\bibitem{Tang97}

H. Li, C. Tang, and N. Wingreen, Phys. Rev. Lett. {\bf 79} (1997) 765.

\bibitem{Marinari}

E. Marinari and G. Parisi, Europhys.Lett. {\bf 19} (1992) 451-458

\bibitem{Irback}

A. Irbäck and F. Potthast,  J. Chem. Phys. {\bf 103} (1995) 10298-10305

\bibitem{Ferrenberg}

A.M. Ferrenberg and R.H. Swendsen, Phys. Rev. Lett. {\bf 61} (1988) 2635;
A.M. Ferrenberg and R.H. Swendsen, Phys. Rev. Lett. {\bf 63} (1989) 1195. 

\bibitem{Borg}

J. Borg, to be published (2000).

\bibitem{parisi}

M. Mezard, G. Parisi and M. Virasoro, {\it Spin Glasses and beyond}, World Scientific,

New York, 1988)

\bibitem{prion} P. M. Harrison, P Bamborough, V. Daggett, S. B. Prusiner and

F. E. Cohen, Curr. Opin. Struct. Biol. {\bf 7}, 53 (1997)

 

\end{thebibliography}
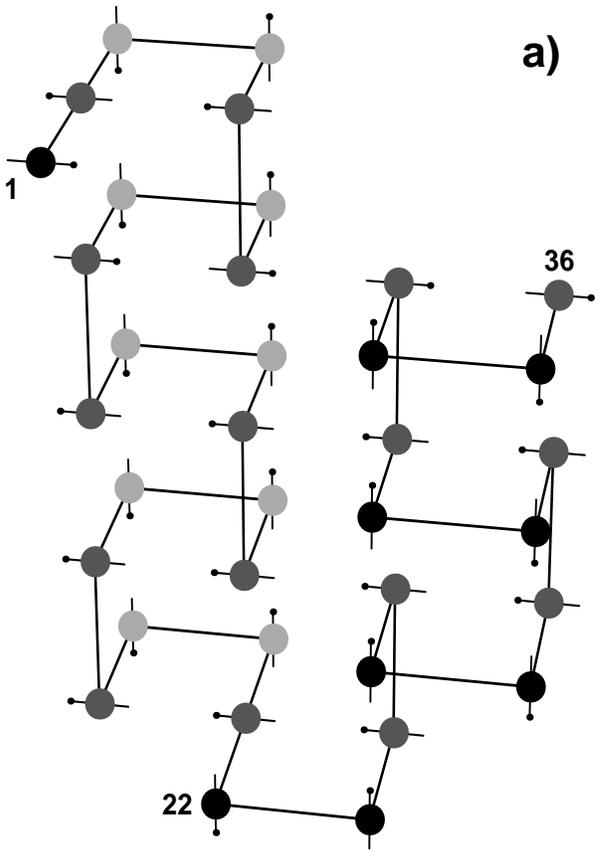
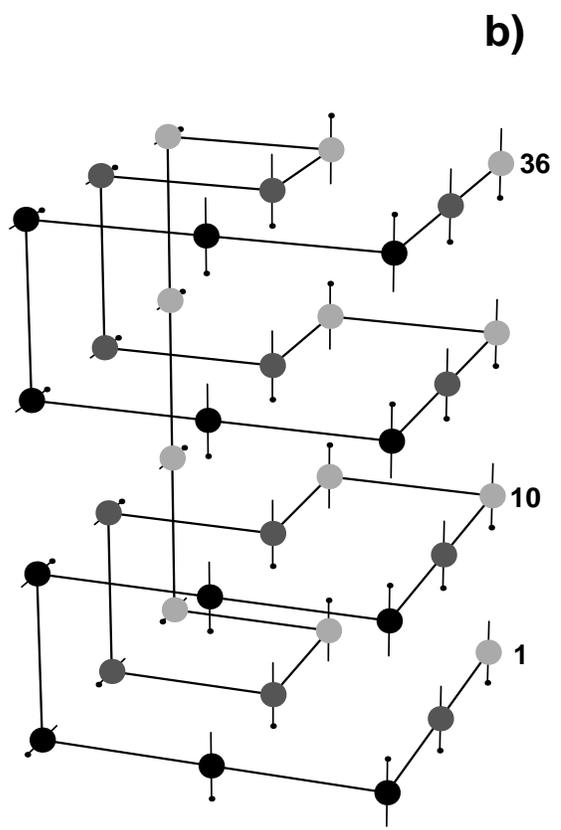
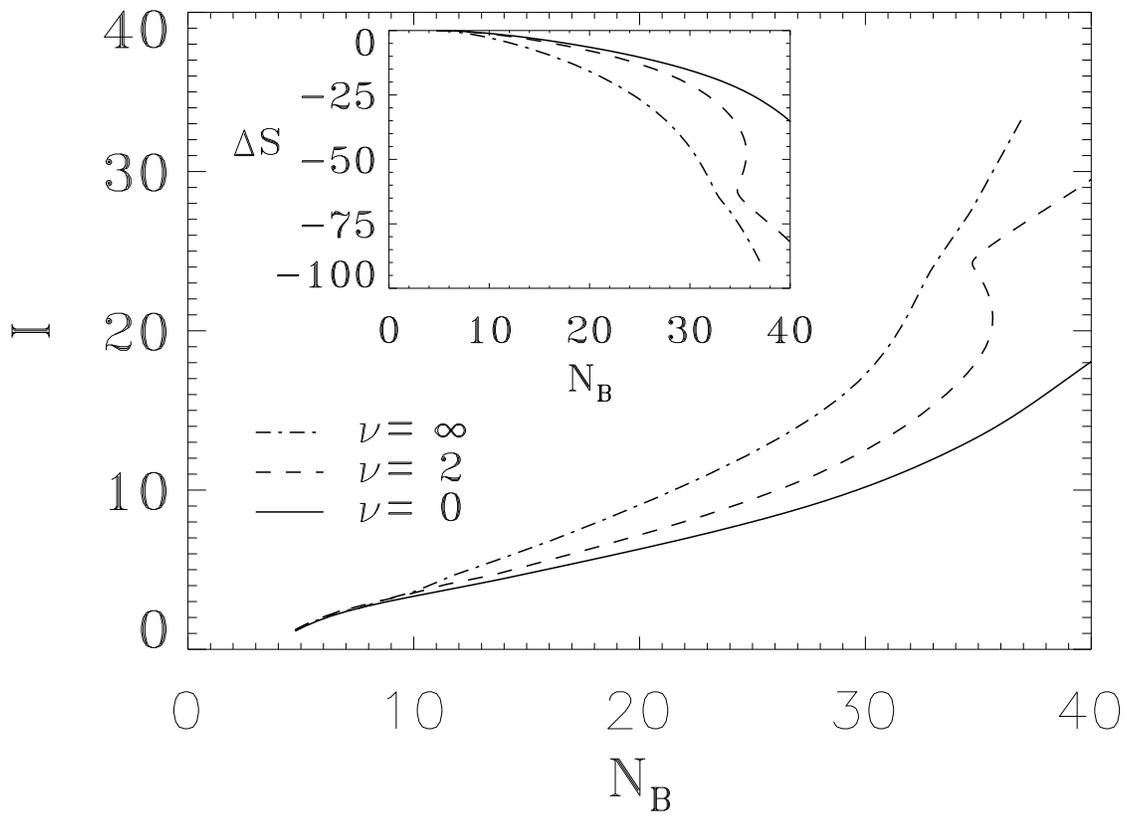
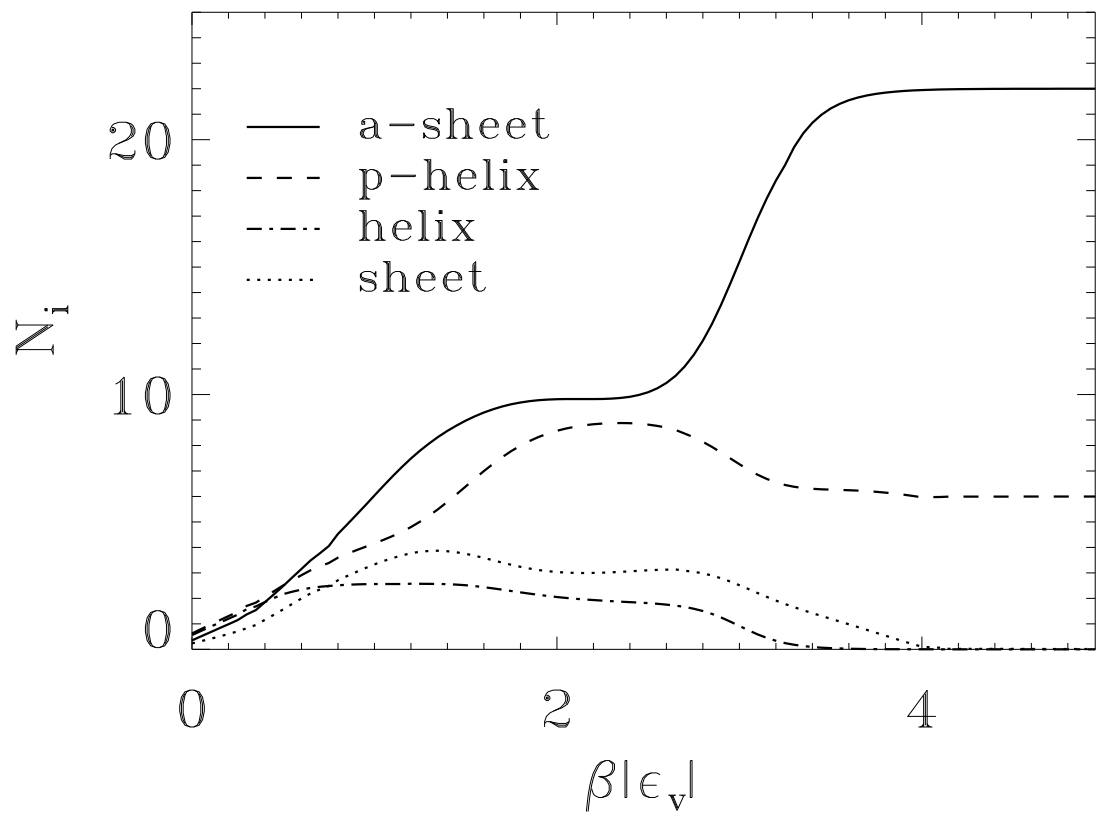
\end{document}